\documentclass[prl,aps,twocolumn]{revtex4-1}

\usepackage[pdftex]{graphicx}
\usepackage{amsbsy,amssymb,amsmath,bm,mathtools}

\usepackage{xspace}
\usepackage{bm}
\usepackage{color}

\usepackage{url}
\usepackage[section]{placeins}
\usepackage[colorlinks=true,linkcolor=blue,citecolor=blue]{hyperref}

\begin{document}


\title{Honeycomb-lattice Gamma model in a magnetic field: \\ hidden N\'eel order and spin-flop transition }

\author{Zhongzheng Tian}

\author{Zhijie Fan}

\author{Preetha Saha}

\author{Gia-Wei Chern}
\affiliation{Department of Physics, University of Virginia, Charlottesville, VA 22904, USA}

\date{\today}
\begin{abstract}
We show that a magnetic field in the high-symmetry direction lifts the macroscopic classical ground-state degeneracy of the honeycomb $\Gamma$ model and induces a long-range magnetic order. While a simple spin-polarized state is stabilized for the ferromagnetic $\Gamma$-exchange, a periodic $\sqrt{3}\times \sqrt{3}$ magnetic order is selected by magnetic field for the antiferromagnetic interaction. We show that the complex spin structure of the tripled unit cell can be described by the magnetization vector and a N\'eel order parameter, similar to those for the spin-flop state of a bipartite antiferromagnet. Indeed, the transition from the low-field plaquette-ordered spin liquid to the field-induced magnetic order can be viewed as a generalized spin-flop transition. An accidental O(2) degeneracy associated with rotation symmetry of the N\'eel vector is broken by either quantum or thermal fluctuations, leaving a six-fold degenerate ground state. At high fields, the breaking of the ground-state $Z_6$ symmetry is through two Berezinskii-Kosterlitz-Thouless transitions that enclose a critical XY phase.  
\end{abstract}

\maketitle

Frustration in physical systems refers to competing interactions that cannot be simultaneously satisfied~\cite{lacroix11,moessner06}. A hallmark of strong frustration is the appearance of accidental degeneracy in which the degenerate states are not related by symmetry operations of the Hamiltonian~\cite{moessner06}. The ground-state degeneracy in some geometrically frustrated magnets, such as kagome or pyrochlore antiferromagnets, scales exponentially with the system size, giving rise to disordered spins even at temperatures well below the exchange energy scale~\cite{chalker92,zhitomirsky08,chern13,moessner98,moessner98b}.  Frustrated interactions also occur in Mott insulators with unquenched orbital degrees of freedom~\cite{kugel82,brink04,tokura00,nussinov15}. A salient feature of such systems is the highly directional orbital exchange interactions, as represented by quantum compass models.   A new type of frustration arises because exchange energy between neighboring pairs along different orientations cannot be simultaneously minimized~\cite{wu08,zhao08,nasu08,rynbach10,chern11}. Similar to geometrically frustrated systems, large accidental degeneracy, sometimes also of macroscopic scale, results from orbital frustration~\cite{chern11,nussinov15}.

Magnetic frustration that involves anisotropic exchange coupling has recently attracted enormous attention. These materials often contain 4$d$ or 5$d$ transition metal elements and are Mott insulators with strong spin-orbit coupling~\cite{pesin10,krempa14}. In these compounds, the localized spin and orbital degrees of freedom are entangled to each other by the relativistic spin-orbit interaction. The resultant composite degree of freedom, which can be viewed as an effective spin variable, preserves the orbital character and is spatially highly anisotropic. Exchange interactions between these local pseudo-spins exhibit strong anisotropy in both real and spin spaces, similar to the orbital-exchange models described above. 

The recent renewed interest in such systems was partly generated by the advent of Kitaev materials~\cite{hermanns18,takagi19,nasu14,trebst17}. Originally proposed as a toy model for fractionalized excitations and topological quantum computing~\cite{kitaev06}, it was later pointed out that Kitaev-type exchange interaction can be realized in $d^5$ transition metal compounds such as $A_2$IrO$_3$ and RuCl$_3$~\cite{jackeli09,chaloupka10,chaloupka13}. The possibility that Kitaev materials might host the elusive quantum spin liquids has generated a flurry of experimental efforts on related compounds and their characterizations. However, other spin-spin interactions, including the isotropic Heisenberg exchange, compete with the Kitaev interaction and often destabilize the spin liquid phase. Considerable efforts have thus been devoted to the study of general anisotropic pseudo-spin interactions in spin-orbit coupled Mott insulators~\cite{rau14,sizyuk14,winter16,janssen17,winter17,stavropoulos19}.

In particular, the anisotropic exchange, also called the $\Gamma$ interaction~\cite{moriya60} is shown to play an important role in compounds such as RuCl$_3$. The $\Gamma$ model on the honeycomb lattice is defined as~\cite{rousochatzakis17,saha19,samarakoon18}
\begin{eqnarray}
	\label{eq:H_Gamma}
	\mathcal{H} = \Gamma \sum_{\gamma} \sum_{\langle ij \rangle \parallel \gamma} (S^\alpha_i S^\beta_j + S^\beta_i S^\alpha_j) - \mathbf H \cdot \sum_i \mathbf S_i,
\end{eqnarray}
where $(\alpha,\beta,\gamma)$ are permutations of $(x, y, z)$. We have also included the Zeeman coupling to a magnetic field $\mathbf H = H\hat{\mathbf n}$ in the $\hat{\mathbf n} \parallel [111]$ direction.  
The honeycomb $\Gamma$~model is a highly frustrated spin system which supports a novel classical spin-liquid ground state~\cite{rousochatzakis17}. 
The  extensive degeneracy associated with the classical ground state is characterized by an emergent global O(3) rotational symmetry and a local $Z_2$ gauge-like symmetry~\cite{rousochatzakis17}.  While the local Ising-gauge symmetry cannot be spontaneously broken~\cite{elitzur75}, the continuous O(3) degeneracy is lifted by quantum or thermal fluctuations~\cite{rousochatzakis17,saha19}. 
Interestingly, the spontaneous breaking of the O(3) symmetry actually corresponds to a breaking of lattice translation symmetry. Through the order-by-disorder mechanism, fluctuations thus induce a sharp phase transition below which an exotic spin liquid with a hidden $\sqrt{3} \times \sqrt{3}$ plaquette order emerges as the semiclassical ground state~\cite{saha19}.

In this paper, we study the effect of magnetic field on the semiclassical honeycomb $\Gamma$ model. The large degeneracy of frustrated magnets renders them susceptible to perturbations brought about by the magnetic field. Indeed, novel field-induced phases such as magnetization plateau and even spin liquid have been reported in both geometrically frustrated magnets~\cite{zhitomirsky02,penc04,zhitomirsky00,seabra16,schulenburg02,fortune09,alicea09,wosnitza16,misguich01,sen08} and Kitaev spin models~\cite{yadav16,Janssen16,chern17,gohlke18,lee20,liu18,zhu18,hickey19,gordon19,patel19,janssen19,balz21}.  In our case, the extensive ground-state degeneracy of the classical $\Gamma$-model is lifted by field along the high-symmetry [111] direction. For the ferromagnetic case with $\Gamma < 0$, the polarized state with spins aligning with the field direction is selected by the field since this particular ferromagnetic state happens to be one of the ground state of the zero-field $\Gamma$ model~\cite{rousochatzakis17}.

\begin{figure}
    \centering
    \includegraphics[width=0.99\columnwidth]{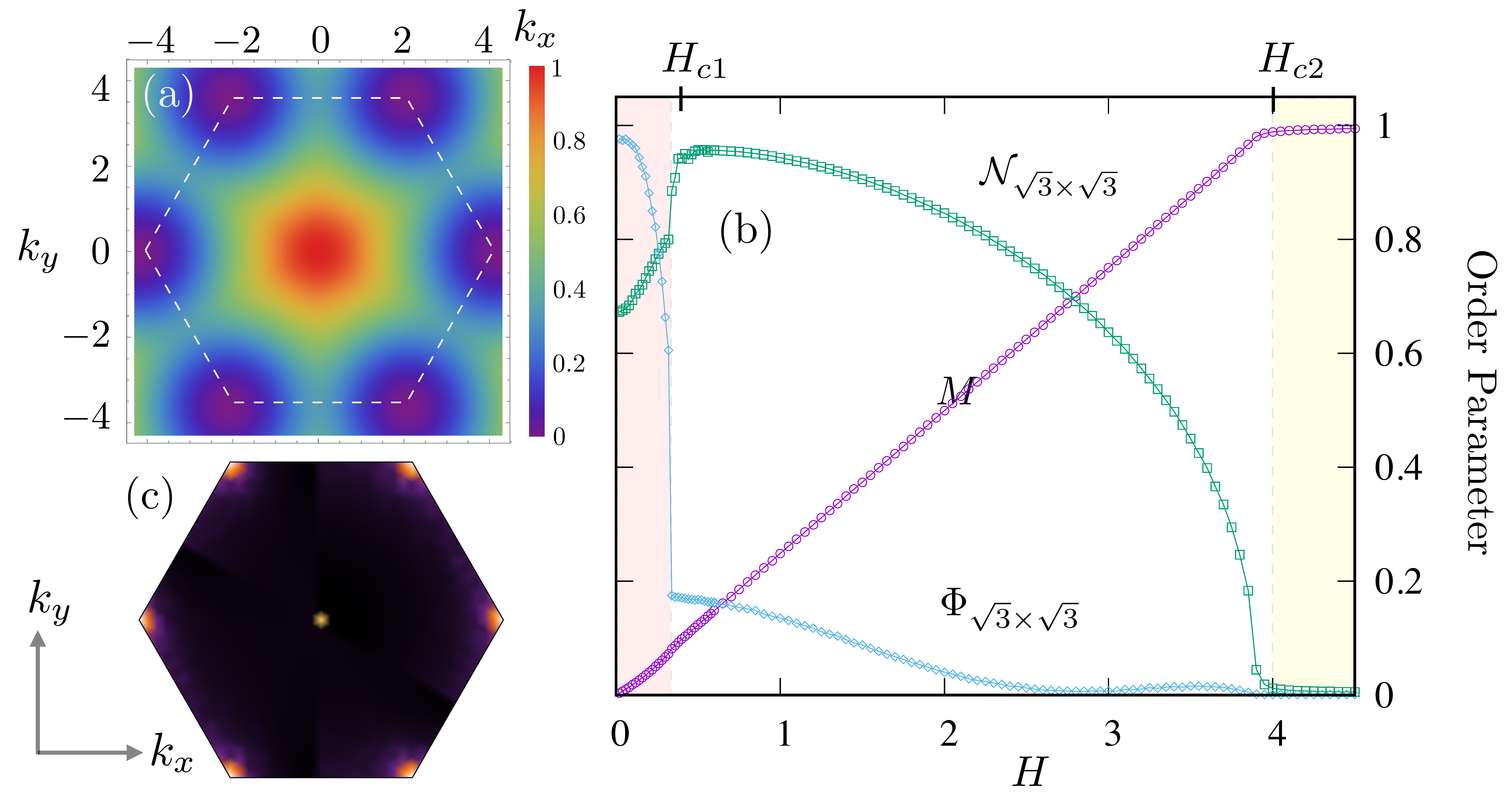}
    \caption{(a) Density plot in $\bm k$-space of the minimum eigen-energy of the fluctuation interaction $E_2$. The dashed line marks the boundary of the first Brillouin zone. (b) Monte Carlo simulation results at temperature $T = 0.01 \Gamma$ showing the field dependence of $\sqrt{3}\times \sqrt{3}$ magnetic order $\mathcal{N}_{\sqrt{3}\times \sqrt{3}}$, plaquette flux order $\Phi_{\sqrt{3}\times \sqrt{3}}$, and magnetization $M$. (c) The static structure factor of the intermediate state from Monte Carlo simulation at $H = 2\Gamma$ and $T = 0.01\Gamma$. }
    \label{fig:m-H}
\end{figure}

The antiferromagnetic $\Gamma$ model, on the other hand, remains frustrated in the presence of magnetic field. To obtain the structure of possible field-induced order, we investigate the stability of the polarized state at large~$H$. To this end, small deviations $\delta\mathbf S_i = \varsigma_i^1 \hat{\mathbf e}_1 + \varsigma_i^2 \hat{\mathbf e}_2$, where $\hat{\mathbf e}_1 = (\hat{\mathbf e}_x + \hat{\mathbf e}_y - 2 \hat{\mathbf e}_z)/\sqrt{6}$ and $\hat{\mathbf e}_2 = (\hat{\mathbf e}_x - \hat{\mathbf e}_y)/\sqrt{2}$ are two unit vectors perpendicular to the field direction, are introduced to the polarized state. Substituting $\mathbf S_i = \sqrt{S^2 - |\delta \mathbf S_i|^2} \, \hat{\mathbf n} + \delta\mathbf S_i$ into Eq.~(\ref{eq:H_Gamma}), and expanding to second order in $\varsigma$, we obtain $\mathcal{H} = E_0 +E_2$, where $E_0 = NS (\Gamma S - H )$ is the energy of the polarized state, $N$ is the number of spins, and $E_2$ denotes a 120$^\circ$-type interaction energy:
\[
	E_2 = \varepsilon \sum_i |\bm\varsigma_i|^2 
	+ \frac{2\Gamma}{3} \sum_\gamma \sum_{\langle ij \rangle \parallel \gamma} \left[ (\bm\varsigma_i \cdot \hat{ \bm \epsilon}_\alpha) (\bm\varsigma_j \cdot \hat{ \bm \epsilon}_\beta) + (i \leftrightarrow j)\right]. \nonumber
\]
Here $\bm\varsigma_i = (\varsigma_i^1, \varsigma_i^2)$, $\varepsilon = H/S - 2 \Gamma$, and $\hat{\bm \epsilon}_x =(\frac{1}{2}, -\frac{\sqrt{3}}{2})$, $\hat{\bm \epsilon}_y = (\frac{1}{2}, \frac{\sqrt{3}}{2})$, $\hat{\bm \epsilon}_z = (-1, 0)$ are three unit vectors that are 120$^\circ$ from each other. This energy can be partially diagonalized using Fourier transform: $E_2 = \sum_{\bm k} \mathbb{U}^*_{\bm k} \cdot \mathbb{H}(\bm k) \cdot \mathbb{U}_{\bm k}$, where $\mathbb{U}_{\bm k} = [ \varsigma^1_{A, \bm k}, \varsigma^2_{A, \bm k}, \varsigma^1_{B, \bm k}, \varsigma^2_{B, \bm k} ]^{\rm t}$, and the subscripts $A$, $B$ denote the two sublattices.  The eigen-mode energy, as a function of momentum $\bm k$, is then given by the eigenvalues of the $4\times 4$ matrix $\mathbb{H}(\bm k)$. 




The lowest eigen-mode energy shown in Fig.~\ref{fig:m-H}(a) exhibits six minima at the corners of the Brillouin zone (BZ), indicating that the most unstable mode has a wave vector $\bm k^* = (\frac{4\pi}{3a}, 0)$, where $a$ is the lattice constant of the honeycomb lattice. This ordering wave vector corresponds to the $\sqrt{3}\times \sqrt{3}$ periodic structure in real space. Consequently, as the magnetic field is lowered below some critical value, the polarized state becomes unstable against the development of the $\sqrt{3}\times \sqrt{3}$ magnetic order. From the analytical solution of the minimum mode energy at the K-point, $\varepsilon_{\bm k^*} = H/S - 4\Gamma$, the instability condition $\varepsilon_{\bm k^*} = 0$ gives an upper critical field $H_{c2} = 4\Gamma S$.

This conclusion is verified by our classical Monte Carlo simulations, for example, the magnetization curve at a low temperature $T = 0.01 \Gamma$, shown in Fig.~\ref{fig:m-H}(b). At small magnetic field, the plaquette-ordered spin liquid remains stable up to some critical field $H_{c1}$, above which the flux order parameter $\Phi_{\sqrt{3}\times \sqrt{3}}$ drops abruptly.  The magnetization $M$ increases linearly in this intermediate regime until spins are fully polarized at $H \gtrsim H_{c2} = 4\Gamma S$. Fig.~\ref{fig:m-H}(c) shows the static structure factor of the intermediate state, which exhibits six peaks at the corners of the BZ in addition to the central peak at $\bm k=0$ due to the field-induced finite magnetization. 

\begin{figure}
    \centering
    \includegraphics[width=0.99\columnwidth]{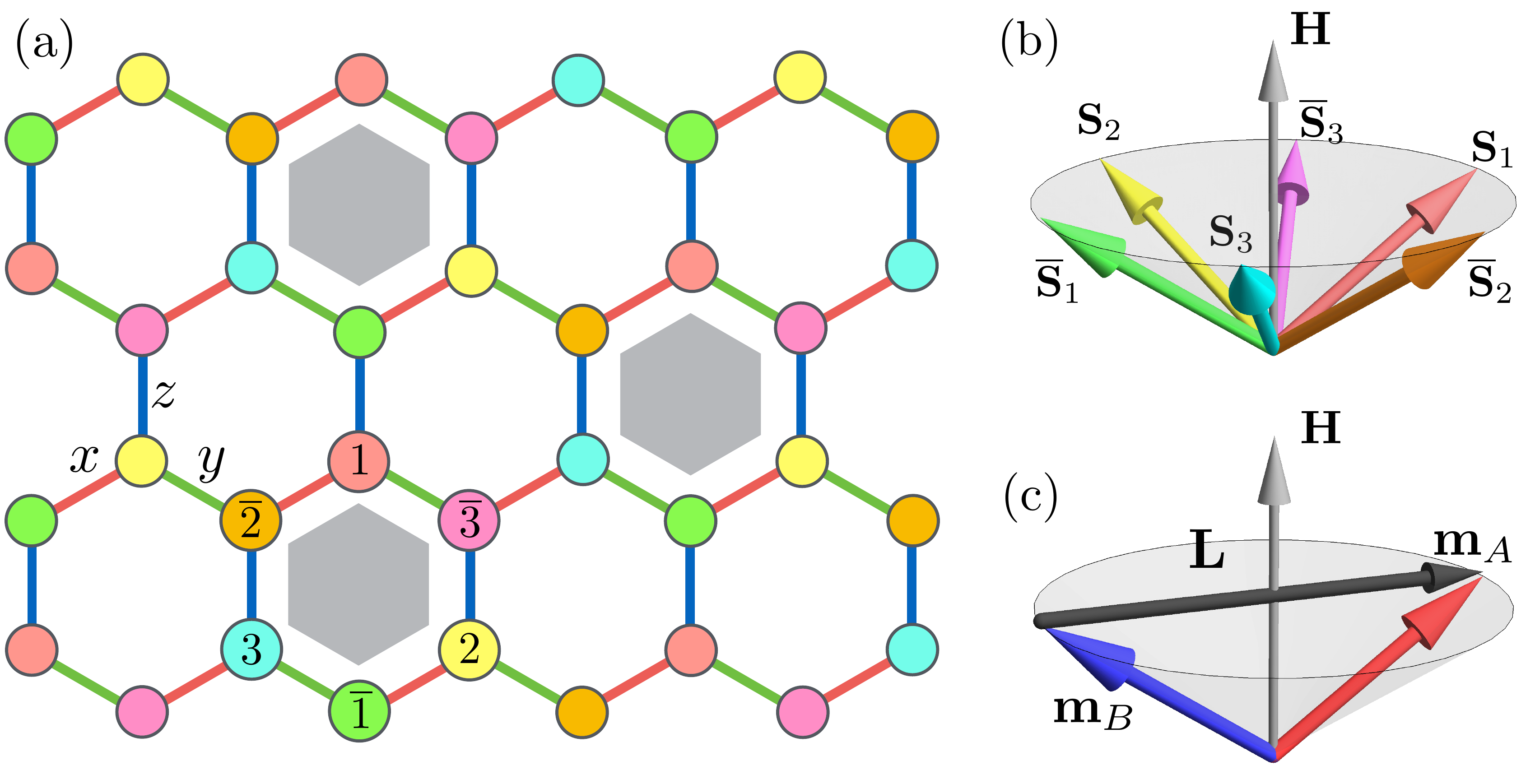}
    \caption{(a) The magnetic ground state of antiferromagnetic $\Gamma$-model in a $[111]$ field. (b) Relative orientation of the six sublattice spins in the  $\sqrt{3}\times \sqrt{3}$ magnetic order shown in panel~(a). The complex spin structure can be described by two sublattice order parameter $\mathbf m_A$ and $\mathbf m_B$ shown in panel~(c).}
    \label{fig:gs}
\end{figure}

We next determine the structure of the $\sqrt{3}\times \sqrt{3}$ state. The six inequivalent spins in the extended unit cell form an 18-dimensional representation of the little group of the K-point. The relevant magnetic order parameters, which can be obtained by examining the irreducible representations, can be very complicated. Our direct energy minimization, however, finds a rather simple magnetic structure which can be described by a N\'eel order parameter. Generalizing parametrization of spins in the special $\sqrt{3}\times \sqrt{3}$ ground state of the zero-field $\Gamma$-model~\cite{saha19}, we introduce two sublattice order parameters $\mathbf m_A = (a, b, c)$ and $\mathbf m_B = (\overline{a}, \overline{ b}, \overline{ c})$. The six sublattice spins of the tripled unit cell, as labeled in Fig.~\ref{fig:gs}, can be expressed as
\begin{eqnarray}
	\label{eq:sublattice-S}
	& & \mathbf S_1=S(a, b, c), \quad \mathbf S_2 = S(b, c, a), \quad \mathbf S_3 = S(c, a, b), \nonumber \\
	& & \overline{\mathbf S}_1 = S(\overline{b}, \overline{a}, \overline{c}), \quad \overline{\mathbf S}_2 = S(\overline{a}, \overline{c}, \overline{b}), 
	\quad \overline{\mathbf S}_3 =S (\overline{c}, \overline{b}, \overline{a}). \quad
\end{eqnarray}
Remarkably, with this parametrization the energy of the complex $\sqrt{3}\times \sqrt{3}$ order is given by
\begin{eqnarray}
	\label{eq:E_gs}
	E/N = \Gamma S^2 \mathbf m_A \cdot \mathbf m_B - \frac{1}{2} H S \hat{\mathbf n} \cdot (\mathbf m_A + \mathbf m_B),
\end{eqnarray}
which is exactly the same as the energy of the spin-flop state of a bipartite antiferromagnet~\cite{chaikin95,fisher74,shapira70}. To obtain the ground state, we introduce the magnetization vector $\mathbf M$ and a ``N\'eel" vector $\mathbf L$ that characterizes the disparity of the two sublattices:
\begin{eqnarray}
	\mathbf M = (\mathbf m_A + \mathbf m_B)/2, \qquad \mathbf L = (\mathbf m_A - \mathbf m_B)/2.
\end{eqnarray}
In the classical ground state, these two order parameters satisfy the conditions: $\mathbf M^2 + \mathbf L^2 = 1$, and $\mathbf M \cdot \mathbf L = 0$. The energy in Eq.~(\ref{eq:E_gs}) is minimized when $\mathbf M$ and $\mathbf L$ are parallel and perpendicular to the field direction, respectively.  The magnetization of the minimum-energy solution is $M = H/4S\Gamma$. The upper critical field obtained from the condition $M = 1$ of fully polarized spins is $H_{c2} = 4\Gamma S$, consistent with that derived from the stability analysis.

\begin{figure}
    \centering
    \includegraphics[width=0.99\columnwidth]{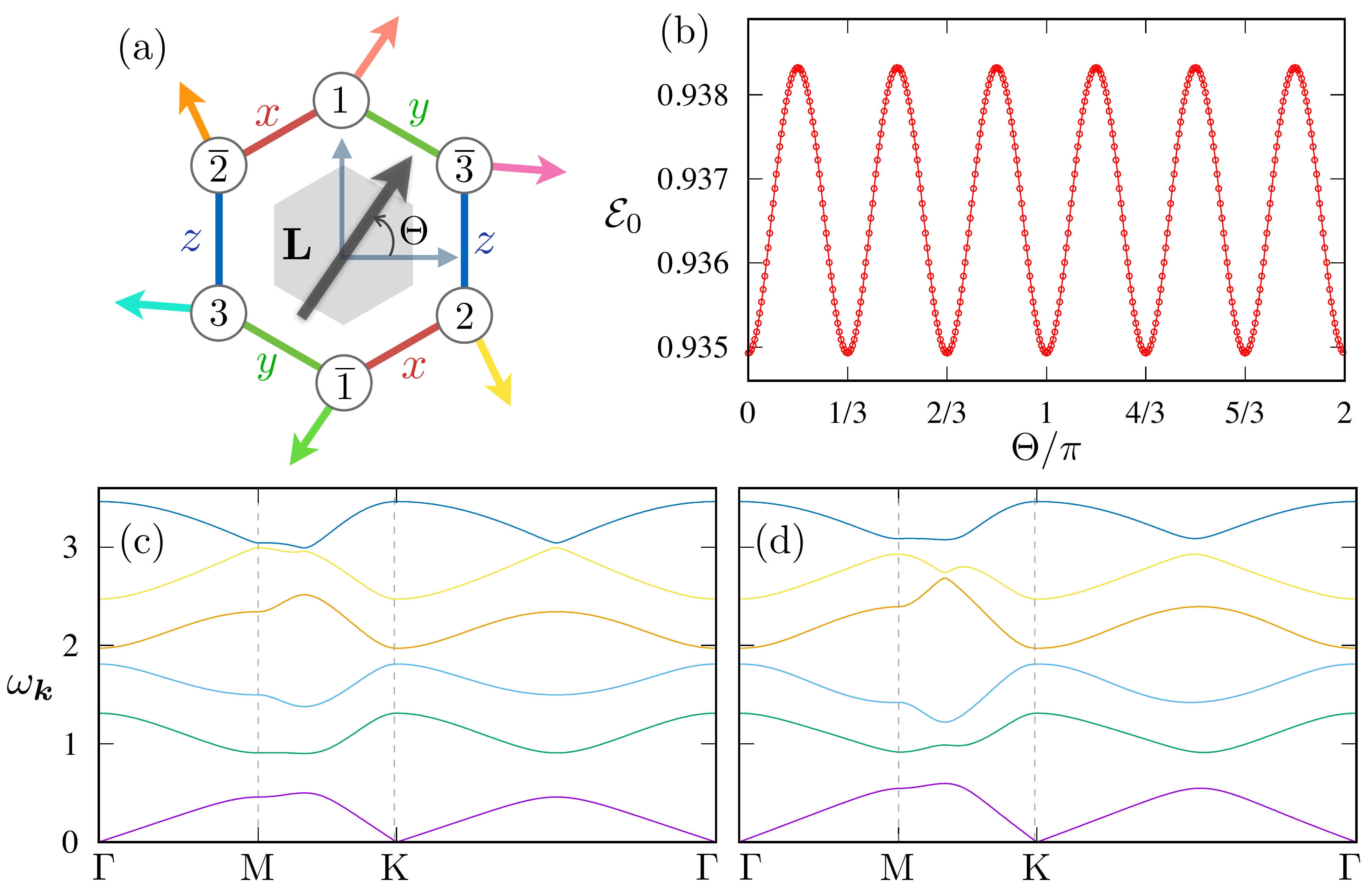}
    \caption{(a) N\'eel vector $\mathbf L = L(\cos\Theta\,\hat{\mathbf e}_1 + \sin\Theta \,\hat{\mathbf e}_2)$ associated with a hexoagon; here $\hat{\mathbf e}_1 = (\hat{\mathbf e}_x + \hat{\mathbf e}_y - 2 \hat{\mathbf e}_z)/\sqrt{6}$ and $\hat{\mathbf e}_2 = (\hat{\mathbf e}_x - \hat{\mathbf e}_y)/\sqrt{2}$ are two unit vectors perpendicular to the [111] field direction. (b) Zero point energy of magnons as a function of angle $\Theta$ for magnetization $M = 0.5$. The spinwave spectra at $\Theta = 0$ and $\pi/6$ are shown in panels~(c) and (d), respectively.}
    \label{fig:e-zero}
\end{figure}

In terms of the order parameters, the energy per spin, $E/N =  \Gamma S^2 (\mathbf M^2 - \mathbf L^2) -  HS \hat{\mathbf n}\cdot \mathbf M$, is invariant under rotation of the N\'eel vector around the field direction. As the $\Gamma$ model itself does not possess such rotation symmetry, this accidental O(2) degeneracy is expected to be lifted when quantum or thermal fluctuations are taken into account. We first consider the quantum order-by-disorder mechanism and outline the linear spinwave calculation for the spin-flop state shown in Fig.~\ref{fig:gs}(b). To this end, we write the N\'eel vector as $\mathbf L = L( \cos\Theta \,\hat{\mathbf e}_1 + \sin\Theta \, \hat{\mathbf e}_2)$, where $L = \sqrt{1 - M^2}$, and $\hat{\mathbf e}_{1, 2}$ are two unit vectors perpendicular to the field direction introduced above; see Fig.~\ref{fig:e-zero}(a). 
In the ground state, the six sublattice spins defined in Eq.~(\ref{eq:sublattice-S}) can be expressed as $\mathbf S_r = S \hat{\bm \eta}^z_{+, r}$ and $\overline{\mathbf S}_r = S \hat{\bm \eta}^z_{-, r}$ ($r = 1, 2, 3$), where the quantization axes are 
\[
	\hat{\bm\eta}^z_{\pm, r} = \pm L \left[ \cos\left( \Theta + \omega_r \right) \hat{\mathbf e}_1 + \sin\left( \Theta +\omega_r \right) \hat{\mathbf e}_2 \right] + M \hat{\mathbf n}.
\]
Here $\pm$ corresponds to A/B sublattice, respectively, and $\omega_r = 0, \frac{2\pi}{3}, \frac{4\pi}{3}$ for $r = 1, 2, 3$, respectively. One can introduce an orthogonal triad of unit vectors for each sublattice by defining $\hat{\bm\eta}^x_{\pm, r} = \mp [\sin(\Theta + \omega_r) \hat{\mathbf e}_1 - \cos(\Theta + \omega_r) \hat{\mathbf e}_2 ]$ and $\hat{\bm \eta}_{\pm, r}^y = \hat{\bm\eta}^z_{\pm, r} \times \hat{\bm\eta}_{\pm, r}^x$. For convenience, we use $K_i = (s_i, r_i)$, where $s_i = \pm 1$ and $r_i = 1, 2, 3$, to denote the magnetic sublattice of site-$i$. Using the Holstein-Primakoff transformation, we write the spin operator at site-$i$ as $\hat{\mathbf S}_i \approx \sqrt{2S} \left( \hat{a}^x_{i} \hat{\bm \eta}_{K_i}^x + \hat{a}^y_{i} \hat{\bm\eta}_{K_i}^y \right) + (S - \hat{a}_i^\dagger \hat{a}^{\,}_i) \hat{\bm\eta}^z_{K_i}$, where $\hat{a}^x_i = (\hat{a}^{\,}_i + \hat{a}^\dagger)/2$, $\hat{a}^y_i = (\hat{a}^{\,}_i - \hat{a}^\dagger_i)/2i$, and $\hat{a}^\dagger_i$ ($\hat{a}^{\,}_i$) are the on-site magnon creation (annihilation) operators. Substituting the $\hat{\mathbf S}_i$ operator into Eq.~(\ref{eq:H_Gamma}), we obtain the following magnon Hamiltonian
\begin{eqnarray}
	\hat{\mathcal{H}} = E_{\rm SF}+ 2 \Gamma S \sum_i  \hat{a}^\dagger_i \hat{a}^{\,}_i  + \sum_{\langle ij \rangle}\sum_{\mu\nu}^{x,y} \hat{a}^\mu_i \mathcal{M}^{\mu\nu}_{ij} \hat{a}^\nu_j, \quad
\end{eqnarray}
where $E_{\rm SF} = -N(\Gamma S^2 + H^2/8\Gamma)$ is the energy of the spin-flop state, the coefficient $\mathcal{M}^{\mu\nu}_{ij} = 2S\hat{\bm\eta}^\mu_{K_i} \cdot {\bm \Gamma}_{ij} \cdot \hat{\bm\eta}^\nu_{K_j}$, and $\bm\Gamma_{ij}$ is the Gamma-interaction matrix on $\langle ij \rangle$ bond.

The magnon Hamiltonian is then diagonalized using Fourier and Bogoliubov transformations. Figs.~\ref{fig:e-zero}(c) and~(d) show the spinwave spectrum $\omega_n(\bm k)$, where $n = 1, \cdots, 6$ is the band index, along high-symmetry directions of the BZ for two different angles $\Theta = 0$ and $\Theta = \pi/6$, respectively. In both cases, a pseudo-Goldstone mode~\cite{rau18} is obtained at the center and corners of the BZ, which can be attributed to the O(2) symmetry combined with a $Z_3$ symmetry associated with the $\sqrt{3}\times \sqrt{3}$ order. The different spectra also means that the quantum zero-point energy, given by the sum $\mathcal{E}_0 = \sum_{n, \bm k} \omega_{n}(\bm k)/2$, depends on the orientation angle $\Theta$ of N\'eel vector. As shown in Fig.~\ref{fig:e-zero}(b), the zero-point energy exhibits six minima at $\Theta = m\pi / 3$, where $m$ is an integer, indicating that these six orientations, related by the hexagonal symmetry of the $\Gamma$-model, are favored by quantum fluctuations. It is worth noting that quantum fluctuations are expected to also gap out the pseudo-Goldstone mode when higher-order magnon interactions are included~\cite{rau18}.

At non-zero but low temperatures the accidental O(2) degeneracy is also lifted by thermal order by disorder, which selects the same six-fold degenerate ground state, as confirmed by our Monte Carlo simulations. However, the O(2) symmetry is restored at further elevated temperatures and persists within a finite window, giving rise to a critical XY phase. Indeed, field-induced XY criticality in the spin-flop state of 2D bipartite antiferromagnets has been reported for both classical and quantum spins~\cite{kosterlitz76,landau81,okwamoto84,pires94,cuccoli03,holtschneider05,leheny99}. As the N\'eel order in the spin-flop state is forced to lie in a plane perpendicular to the field direction, the magnet effectively becomes an XY system.

\begin{figure}[t]
    \centering
    \includegraphics[width=0.99\columnwidth]{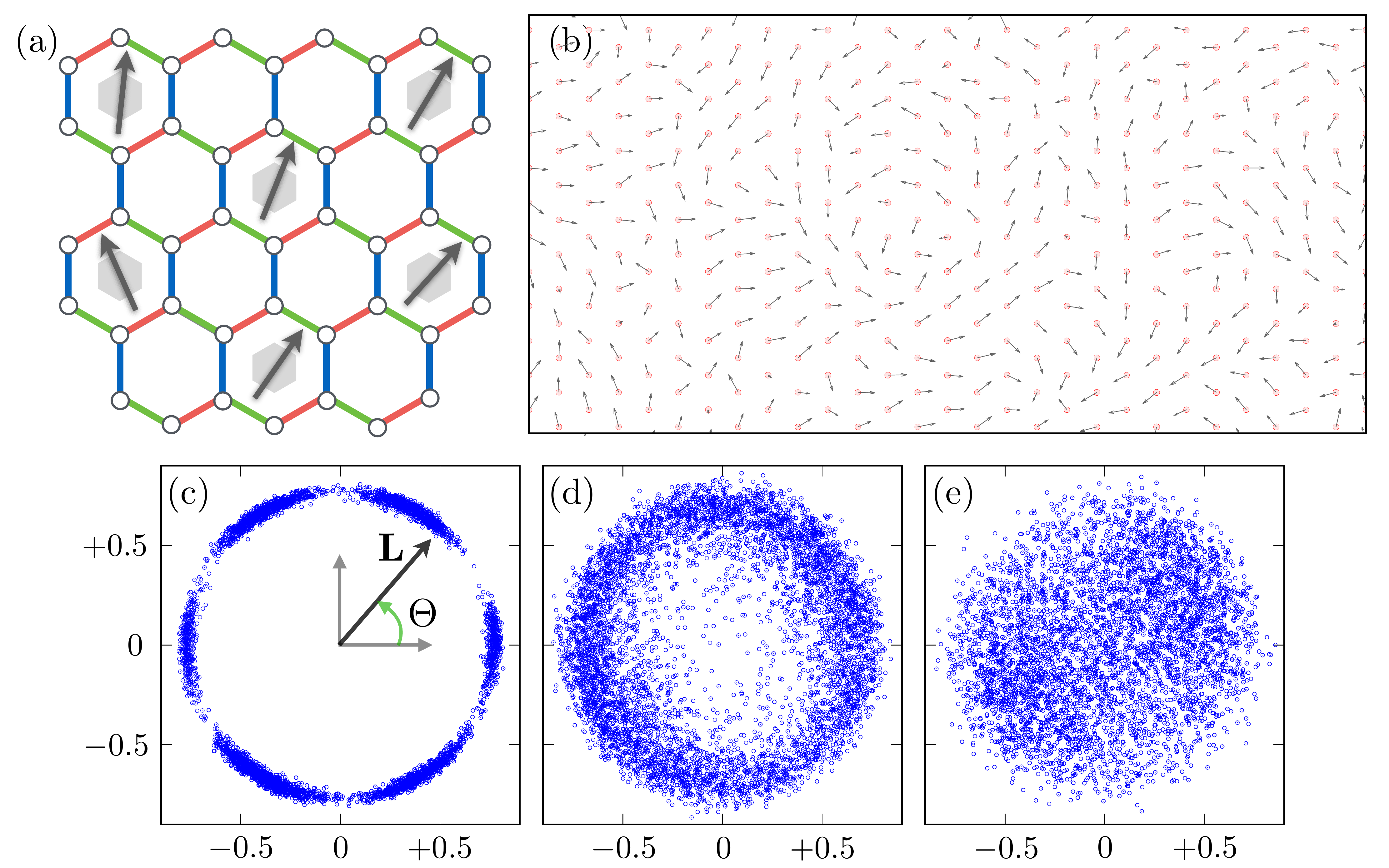}
    \caption{(a) Schematic diagram showing ordering of local N\'eel vectors $\mathbf L_\alpha \sim (\cos\Theta_\alpha, \sin\Theta_\alpha)$ associated with shaded hexagons representing a local magnetic unit cell. (b) A snapshot of $\mathbf L$-vector at $H = 2 \Gamma$ and $T = 0.1\Gamma$. The histogram of the $\mathbf L$ vectors at temperatures (c) $T = 0.01\Gamma$, (d) $T = 0.1\Gamma$, and (e) $T = 0.2\Gamma$. }
    \label{fig:L-hist}
\end{figure}

For the Gamma model, a local N\'eel vector can be defined for each hexagon, for example $L^x = S^x_1 - \overline{S}^y_1 + S^z_2 - \overline{S}^x_2 + S^y_3 - \overline{S}^z_3$, and so on for the $y$ and $z$ components; see Eq.~(\ref{eq:sublattice-S}). These local N\'eel vectors behave as XY spins at low temperatures in the spin-flop state of the $\Gamma$ model. A snapshot of the hexagonal $\mathbf L$-vectors at $T = 0.1\Gamma$ is shown in Fig.~\ref{fig:L-hist}(b). Importantly, the low-temperature behaviors of the Gamma model can be described by a ferromagnetic XY model subject to a six-state clock anisotropy. 
The breaking of the $Z_6$ symmetry in this model is known to go through two Berezinskii-Kosterlitz-Thouless (BKT) transitions which enclose an intermediate critical XY phase~\cite{cardy80,challa86}, a scenario that is confirmed in our Monte Carlo simulations. As demonstrated by the histogram of local N\'eel vectors at three different temperatures shown in Fig.~\ref{fig:L-hist}(c)--(e), an O(2) rotational symmetry emerges in the intermediate critical phase where the spin-spin correlation decays algebraically with distance.

\begin{figure}[t]
\includegraphics[width=0.95\columnwidth]{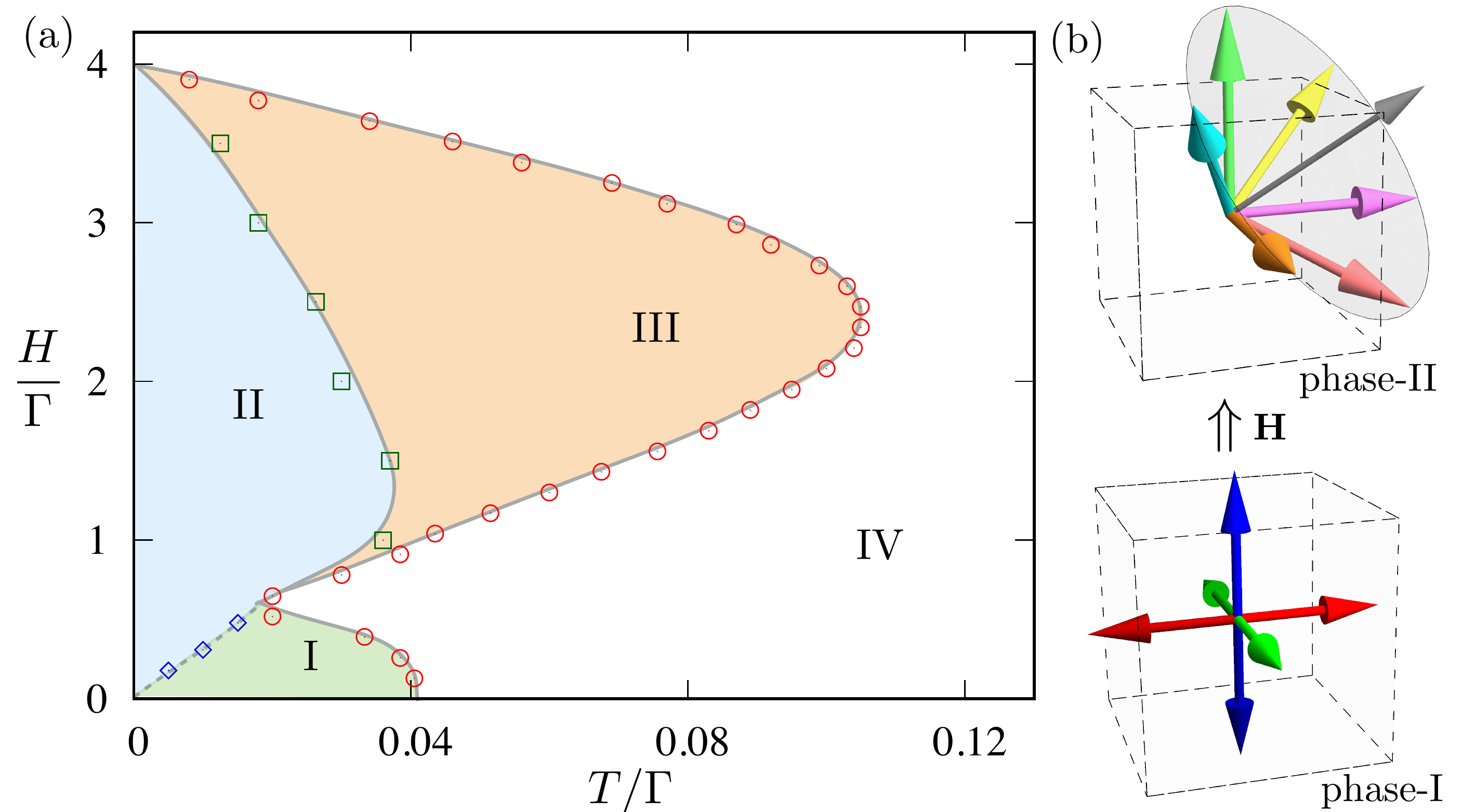}
\caption{(a) Schematic phase diagram of the classical $\Gamma$ model obtained from Monte Carlo simulations. The various phases are: (I) spin liquid with hexagonal flux order that breaks the lattice translation symmetry. (II) Long-range $\sqrt{3}\times \sqrt{3}$ magnetic order with a tripled unit cell. (III) Critical XY phase with an emergent O(2) symmetry of N\'eel vectors. (IV) Classical spin liquid with short-ranged correlation that is smoothly connected to high-$T$ paramagnet and the polarized state at high field. (b) the transition from the plaquette spin liquid to the $\sqrt{3}\times \sqrt{3}$ magnetic order resembles a spin-flop transiton in bipartite antiferromagnet with weak anisotropy. }
    \label{fig:phases}
\end{figure}

The various thermodynamic phases obtained from Monte Carlo simulations are summarized in the magnetic field $H$ versus temperature $T$ phase diagram shown in Fig.~\ref{fig:phases}(a). Depending on strength of the magnetic field, the $\Gamma$ model follows two different routes to reach the $\sqrt{3}\times \sqrt{3}$ magnetic order. While the high-field scenario is through two BKT transitions described above, at small magnetic fields, the system undergoes a crossover and two phase transitions to reach the $\sqrt{3}\times \sqrt{3}$ magnetic ground state. As temperature is lowered below the exchange energy scale, the magnet first enters a classical spin liquid regime with short-range correlation~\cite{rousochatzakis17}. This is followed by another spin liquid with $\sqrt{3}\times \sqrt{3}$ ordering of plaquette fluxes through a continuous phase transition~\cite{saha19}. Upon further lowering the temperature, the plaquette spin liquid phase stabilized by its configurational entropy gives way to the energetically favored magnetic ground state via a first-order transition. 

Despite both having the same wave vector, the ordering of hexagonal fluxes is incompatible with that of spins, which is why the transition between them is of first-order. It is also instructive to understand this discontinuous transition from the viewpoint of spin orientations. The ordering of the fluxes is accompanied by the alignment of spins toward the cubic $x, y, z$ directions due to order-by-disorder~\cite{rousochatzakis17,saha19}. Notably, the N\'eel order parameter can still be used to describe the ``opposite" orientations of the two sublattices in the antiferromagnetic case, although {\em no} long-rang spin order develops because of the emergent Ising pseudo-gauge symmetry. The order-by-disorder mechanism at small fields thus effectively induces a cubic anisotropy for the N\'eel vector: $E_{\rm cubic} = -D( L_x^4 + L_y^4 + L_z^4)$. The competition between this anisotropy and the zeeman coupling to magnetic field leads to a first-order transition similar to the well-studied spin-flop transition~\cite{chaikin95}; see Fig.~\ref{fig:phases}(b).

To summarize, we have uncovered a novel field-induced magnetic ground state in the antiferromagnetic honeycomb Gamma model. This complex magnetic order with a tripled unit cell is a spin-flop state in disguise, and can be described by a hidden N\'eel order parameter. Moreover, we show that the first-order transition between the low-field spin liquid with an effective cubic spin anisotropy and the high-field magnetic order resembles the spin-flop transition in a bipartite antiferromagnet.  Although our semiclassical analysis only applies to large-spin Gamma model, it is likely that this magnetic order is stabilized at high magnetic field even for quantum spin-1/2. A related intriguing question is what happens to the ground state of spin-1/2 Gamma model, which seems to be a gapless spin liquid that is proximate to a zigzag order~\cite{wang19,luo21}, in the presence of magnetic field. Also of interest is the effect of other exchange interactions on the spin-flop state of the $\Gamma$ model reported here. Our work sheds a new light on the nature of complex magnetic structures in such frustrated spin-orbit systems.

This work is partially supported by the Center for Materials Theory as a part of the Computational Materials Science (CMS) program, funded by the US Department of Energy, Office of Science, Basic Energy Sciences, Materials Sciences and Engineering Division.
The authors also acknowledge the support of Advanced Research Computing Services at the University of Virginia.

\end{document}